\documentclass[runningheads]{cl2emult}

\usepackage{makeidx}  % allows index generation
\usepackage{graphicx} % standard LaTeX graphics tool
\usepackage{subeqnar} % subnumbers individual equations
\usepackage{multicol} % used for the two-column index
\usepackage{cropmark} % cropmarks for pages without
\usepackage{eso}      % placeholder for figures
\makeindex            % used for the subject index
\newcommand{\boldthing}[1]{\mbox{\bf #1}}
\newcommand{\myindex}[1]{\mbox{\scriptsize #1}}

\newcommand{\mychoose}[2]{{#1}\choose{#2}}

\newcommand{\bigx}{\boldthing{X}}
\newcommand{\boldx}{\boldthing{x}}
\newcommand{\boldxsub}[1]{{\boldx}_{#1}}
\newcommand{\boldxi}{\boldxsub{i}}
\newcommand{\boldxj}{\boldxsub{j}}
\newcommand{\boldy}{\boldthing{y}}

\newcommand{\countvar}{C}
\newcommand{\countsub}[1]{{\countvar}_{\myindex{#1}}}
\newcommand{\cleft}{\countsub{left}}
\newcommand{\cright}{\countsub{right}}
\newcommand{\limitsub}[2]{{\mbox{\it #1}}_{#2}}

\newcommand{\thingsep}[1]{\limitsub{s}{#1}}

\newcommand{\hisep}{\thingsep{hi}}
\newcommand{\kdnode}{n}
\newcommand{\thingfield}[2]{{#1}.{\myindex{\sc #2}}}
\newcommand{\kdtree}{{\it k}d-tree}
\newcommand{\kdtrees}{{\it k}d-trees}

\newcommand{\losep}{\thingsep{lo}}
\newcommand{\varname}[1]{\mbox{\sc #1}}
\newcommand{\maxdist}{\varname{MaxDist}}
\newcommand{\mindist}{\varname{MinDist}}
\newcommand{\mbw}{\varname{MinBoxWidth}}
\newcommand{\pointset}{S}
\newcommand{\query}{\boldthing{q}}
\newcommand{\algname}[1]{\mbox{\bf #1}}
\newcommand{\rangecount}{\algname{RangeCount}}
\newcommand{\rangecountbetweenseps}{\algname{RangeCountBetweenSeparations}}
\newcommand{\rangesearch}{\algname{RangeSearch}}
\newcommand{\rmin}{r_{\myindex{min}}}
\newcommand{\singletreetwopoint}{\algname{SingleTree2Point}}
\newcommand{\sleft}{{\pointset}_{\myindex{left}}}
\newcommand{\sright}{{\pointset}_{\myindex{right}}}

\newcommand{\thinghr}[1]{\thingfield{#1}{BoundBox}}
\newcommand{\thingleft}[1]{\thingfield{#1}{Left}}
\newcommand{\thingnumpoints}[1]{\thingfield{#1}{NumPoints}}
\newcommand{\thingright}[1]{\thingfield{#1}{Right}}
\newcommand{\thingsplitdim}[1]{\thingfield{#1}{SplitDim}}
\newcommand{\thingsplitval}[1]{\thingfield{#1}{SplitValue}}

\newcommand{\kdnodehr}{\thinghr{\kdnode}}
\newcommand{\kdnodeleft}{\thingleft{\kdnode}}
\newcommand{\kdnodenumpoints}{\thingnumpoints{\kdnode}}
\newcommand{\kdnoderight}{\thingright{\kdnode}}
\newcommand{\kdnodesplitdim}{\thingsplitdim{\kdnode}}
\newcommand{\kdnodesplitval}{\thingsplitval{\kdnode}}

\newcommand{\kdnodesub}[1]{{\kdnode}_{\myindex{#1}}}
\newcommand{\kdnodea}{\kdnodesub{a}}
\newcommand{\kdnodeb}{\kdnodesub{b}}
\newcommand{\dualtreecount}{\algname{DualTreeCount}}
\newcommand{\npt}{\algname{FastNPoint}}

\newcommand{\kdnodesubhr}[1]{\thinghr{\kdnodesub{#1}}}

\newcommand{\kdnodesubnumpoints}[1]{\thingnumpoints{\kdnodesub{#1}}}

\newcommand{\kdnodeahr}{\kdnodesubhr{a}}
\newcommand{\kdnodebhr}{\kdnodesubhr{b}}
\newcommand{\kdnodeanumpoints}{\kdnodesubnumpoints{a}}
\newcommand{\kdnodebnumpoints}{\kdnodesubnumpoints{b}}
\newcommand{\kdnodeihr}{\kdnodesubhr{i}}
\newcommand{\kdnodejhr}{\kdnodesubhr{j}}
\newcommand{\kdnodeinumpoints}{\kdnodesubnumpoints{i}}

\newcommand{\kdnodetosplit}{{\kdnode}^{*}}
\newcommand{\kdnodenonsplit}{{\kdnode}^{-}}
\newcommand{\kdnodesplitleft}{\thingleft{\kdnodetosplit}}
\newcommand{\kdnodesplitright}{\thingright{\kdnodetosplit}}

\newcommand{\knroot}{\kdnodesub{root}}
\newcommand{\lomat}{L}
\newcommand{\himat}{H}
\newcommand{\indicator}[3]{I(#1,#2,#3)}
\newcommand{\lomatsub}[2]{{\lomat}_{(#1,#2)}}
\newcommand{\himatsub}[2]{{\himat}_{(#1,#2)}}
\newcommand{\boldxisub}[1]{\boldxsub{i_{#1}}}

\newcommand{\allsubsumed}{\varname{AllSubsumed}}
\newcommand{\assign}{\mbox{\tt :=}}
\newcommand{\true}{\varname{TRUE}}
\newcommand{\false}{\varname{FALSE}}
\newcommand{\istar}{i^{*}}

\newcommand{\kdnodeistarleft}{\thingleft{\kdnodesub{$\istar$}}}
\newcommand{\kdnodeistarright}{\thingright{\kdnodesub{$\istar$}}}
\newcommand{\threshntuples}{T}
\newcommand{\maxeffect}{\countsub{max}}
\newcommand{\countlo}{\countsub{lo}}
\newcommand{\counthi}{\countsub{hi}}
\newcommand{\countapprox}{\countsub{approx}}

\begin{document}
\title*{Fast Algorithms and Efficient Statistics: N--point Correlation Functions}
\toctitle{Fast Algorithms and Efficient Statistics: N--point Correlation Functions}
\titlerunning{Fast Algorithms and Efficient Statistics: N--point Correlation Functions}                                                       
\author{Andrew Moore\inst{1}
\and Andy Connolly\inst{2}
\and Chris Genovese\inst{1}
\and Alex Gray\inst{1}
\and Larry Grone\inst{1}
\and Nick Kanidoris II\inst{1}
\and Robert Nichol\inst{1}
\and Jeff Schneider\inst{1}
\and Alex Szalay\inst{3}
\and Istvan Szapudi\inst{4}
\and Larry Wasserman\inst{1}
}
\authorrunning{Moore et al.}
\institute{Carnegie Mellon Univ., 5000 Forbes Ave., Pittsburgh, PA-15217
\and Dept. of Physics and Astronomy, Univ. of Pittsburgh, Pittsburgh, PA-15260 
\and Dept. of Physics and Astronomy, Johns Hopkins Univ., Baltimore, MD-21218 
\and CITA, Univ. of Toronto, Toronto, Ontario, M5S 3H8, Canada
}
\maketitle              

\begin{abstract}
\index{abstract} We present here a new algorithm for the fast computation of
$N$--point correlation functions in large astronomical data sets. The algorithm
is based on {\kdtrees} which are decorated with cached sufficient statistics
thus allowing for orders of magnitude speed--ups over the naive non-tree-based
implementation of correlation functions. We further discuss the use of
controlled approximations within the computation which allows for further
acceleration.  In summary, our algorithm now makes it possible to compute
exact, all--pairs, measurements of the 2, 3 and 4--point correlation functions
for cosmological data sets like the Sloan Digital Sky Survey (SDSS; York et
al. 2000) and the next generation of Cosmic Microwave Background experiments
(see Szapudi et al. 2000).

\end{abstract}

\section{Introduction}
Correlation functions are some of the most widely used statistics within
astrophysics (see Peebles 1980 for a extensive review). They are often used to
quantify the clustering of objects in the universe ({\it e.g.} galaxies,
quasars {\it etc.}) compared to a pure Poission process.  More recently, they
have also been used to measure fluctuations in the Cosmic Microwave Background
(see Szapudi et al. 2000).  On large scales, the higher--order correlation
functions (3-point and above) can be used to test several fundamental
assumptions about the universe; for example, our hierarchical scenario for
structure formation, the Gaussianity of the initial conditions as well as
testing various models for the biasing between the luminous and dark
matter. The reader is referred to Szapudi (2000), Szapudi et al. (1999a,b) and
Scoccimarro (2000; and references therein) for an overview of the usefulness
of $N$--point correlation functions in constraining cosmological models.

Over the coming decade, several new, massive cosmological surveys will become
available to the astronomical community.  In this new era, the quality and
quantity of data will warrant a more sophisticated analysis of the
higher--order correlation functions of galaxies (and other objects) over the
largest range of scales possible.  Our ability to perform such studies will be
severely limited by the computational time needed to compute such functions
and no-longer by the amount of data available. In this paper, we address this
computational ``bottle-neck'' by outlining a new algorithm that uses
innovative computer science to accelerate the computation of $N$--point
correlation functions far beyond the naive $O(R^N)$ scaling law (where $R$ is
the number of objects in the dataset and $N$ is the power of correlation
function desired).

The algorithm presented here was developed as part of the ``Computational
AstroStatistics'' collaboration (see Nichol et al. 2000) and is a member of a
family of algorithms for a very general class of statistical computations,
including nearest-neighbor methods, kernel density estimation, and clustering.
The work presented here was initially presented by Gray \& Moore (2001) and
will soon be discussed in a more substantial paper by Connolly et al.
(2001). In this conference proceeding, we provide a brief review of {\kdtrees}
(Section \ref{kdtr}), a discussion of the use of {\kdtrees} in range searches
(Section \ref{range}), an overview of the development of a fast $N$--point
correlation function code (Section \ref{npt}) as well as presenting the
concept of controlled approximations in the calculation of the correlation
function (Section \ref{apprx}). In Section 6, we provide preliminary results
on the computation speed-up achieved with this algorithm and discuss future
prospects for further advances in this field through the use of other tree
structures. 

\section{Review of {\kdtrees}}
\label{kdtr}

Our fast $N$--point correlation function algorithm is built upon the {\kdtree}
data structure which was introduced by Friedman et al. (1977). A {\kdtree}
is a way of organizing a set of datapoints in $k$-dimensional space in such a
way that once built, whenever a query arrives requesting a list all points in
a neighborhood, the query can be answered quickly without needing to scan
every single point.

The root node of the {\kdtree} owns all the data points.  Each non-leaf-node
has two children, defined by a splitting dimension $\kdnodesplitdim$ and a
splitting value $\kdnodesplitval$. The two children divide their parent's data
points between them, with the left child owning those data points that are
strictly less than the splitting value in the splitting dimension, and the
right child owning the remainder of the parent's data points:
\begin{eqnarray}
\boldxi \in \kdnodeleft & \Leftrightarrow &
\boldxi[\kdnodesplitdim] < \kdnodesplitval \mbox{ and } 
\boldxi \in \kdnode \\
\boldxi \in \kdnoderight & \Leftrightarrow &
\boldxi[\kdnodesplitdim] \geq \kdnodesplitval \mbox{ and } 
\boldxi \in \kdnode
\end{eqnarray}
As an example, some of the nodes of a {\kdtree} are illustrated in
Figures~1.

\begin{figure}[t]
\includegraphics[width=1.0\textwidth]{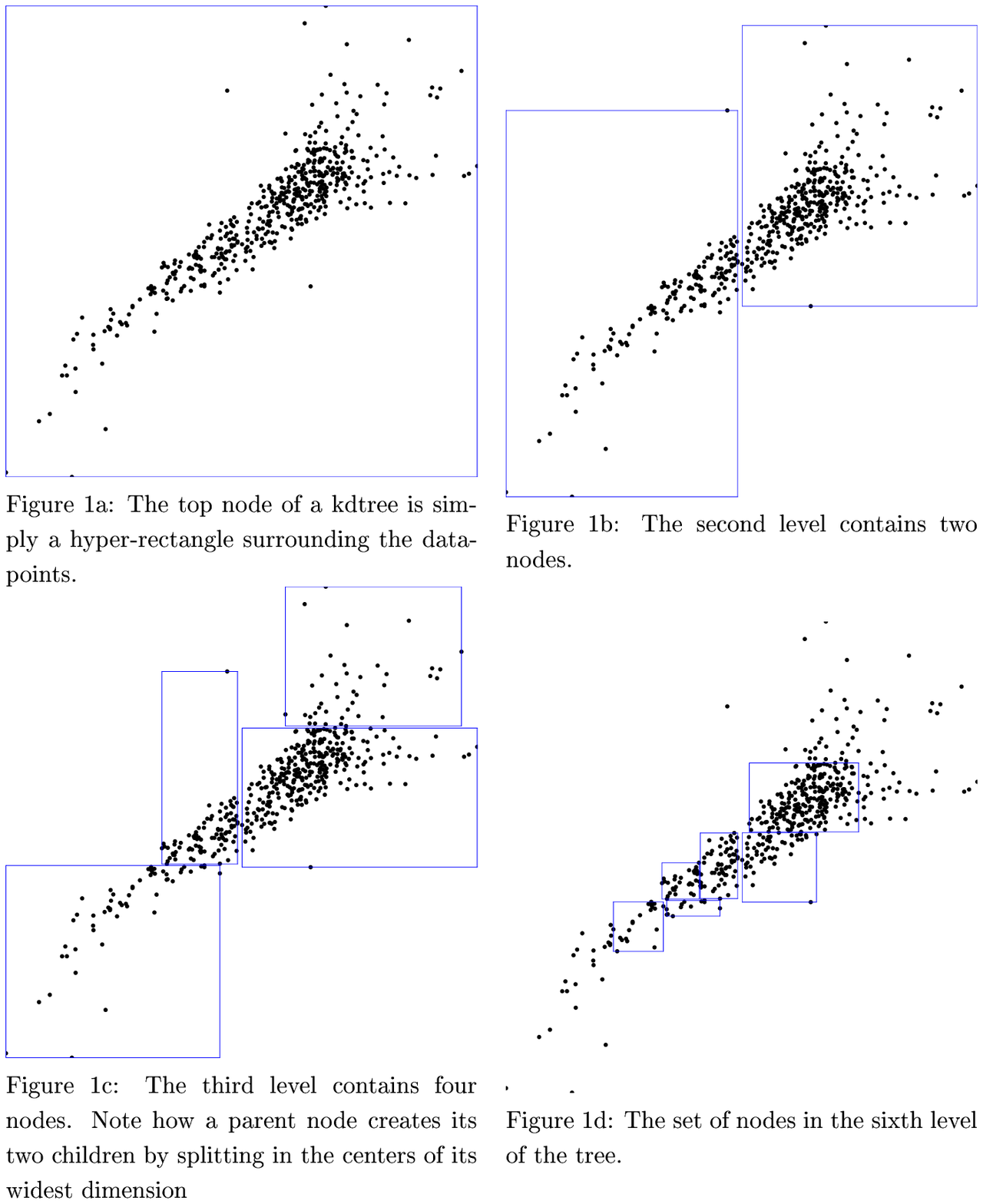}
\end{figure}

{\kdtrees} are usually constructed top-down, beginning with the full set of
points and then splitting in the center of the widest dimension. This produces
two child nodes, each with a distinct set of points. This procedure is then
repeated recursively on each of the two child nodes.

A node is declared to be a leaf, and is left unsplit, if the widest dimension
of its bounding box is $\leq$ some threshold, $\mbw$. A node is also left
unsplit if it denotes fewer than some threshold number of points, $\rmin$. A
leaf node has no children, but instead contains a list of $k$-dimensional
vectors: the actual datapoints contained in that leaf.  The values $\mbw = 0$
and $\rmin = 1$ would cause the largest {\kdtree} structure because all leaf
nodes would denote singleton or coincident points. In practice, we set $\mbw$
to $1\%$ of the range of the data point components and $\rmin$ to around
10. The tree size and construction thus cost considerably less than these
bounds because in dense regions, tiny leaf nodes are able to summarize dozens
of data points. The operations needed in tree-building are computationally
trivial and therefore, the overhead in constructing the tree is negligible.
Also, once a tree is built it can be re-used for many different analysis
operations.

Since the introduction of {\kdtrees}, many variations of them have
been proposed and used with great success in areas such as
databases and computational geometry (Preparata \& Shamos 1985).
R-trees (Guttman 1984) are designed for
disk resident data sets and efficient incremental addition of data.
Metric trees (see Uhlmann 1991) place hyperspheres around tree
nodes, instead of axis-aligned splitting planes. In all cases, the
algorithms we discuss in this paper could be applied equally
effectively with these other structures. For example, Moore (2000)
shows the use of metric trees for accelerating several clustering and
pairwise comparision algorithms.

\section{Range Searching}
\label{range}
Before proceeding to fast $N$--point calculations, we will begin with a very
standard {\kdtree} search algorithm that could be used as a building block for
fast 2-point computations.

For simplicity of exposition we will assume the every node of the {\kdtree}
contains one extra piece of information: the bounding box of all the points it
contains. Call this box $\kdnodehr$. The implication of this is that every
node must contain two new $k$ dimensional vectors to represent the lower and
upper limits of each dimension of the bounding box.  The range search
operation takes two inputs. The first is a $k$-dimensional vector $\query$
called the {\em query point}. The second is a separation distance
$\hisep$. The operation returns the complete set of points in the {\kdtree}
that lie within distance $\hisep$ of $\query$.

\begin{itemize}
\item
$\rangesearch(\kdnode,\query,\hisep)$\\
Returns a set of points $\pointset$ such that
\begin{equation}
\boldx \in \pointset \Leftrightarrow \boldx \in \kdnode {\mbox{ and }}
| \boldx - \query | \leq \hisep
\end{equation}
\item
Let $\mindist$ := the closest distance from $\query$ to $\kdnodehr$.
\item
If $\mindist \geq \hisep$ then it is impossible that any point in
$\kdnode$ can be within range of the query. So simply return the
empty set of points without doing any further work.
\item
Else, if $\kdnode$ is a leaf node, we must iterate through all the
datapoints in its leaf list. For each point, find if it is within distance
$\hisep$ of $\query$. If so, add it to the set of results.
\item
Else, $\kdnode$ is not a leaf node. Then:
\begin{itemize}
\item
Let $\sleft := \rangesearch(\kdnodeleft,\query,\hisep)$
\item
Let $\sright := \rangesearch(\kdnoderight,\query,\hisep)$
\item
Return $\sleft \cup \sright$.
\end{itemize}
\end{itemize}

Figure 2a shows the result of running this algorithm in two dimensions. Many
large nodes are pruned from the search. 117 distance calculations were needed
for performing this range search, compared with 499 that would have been
needed by a naive method.

Note that it is not essential that {\kdtree} nodes have bounding boxes
explicitly stored. Instead a hyper-rectangle can be passed to each recursive
call of the above function and dynamically updated as the tree is searched.

Range searching with a {\kdtree} can be much faster than without if the
range is small, containing only a small fraction of the total number of 
datapoints. But what if the range is large? Figure 2b shows an
example in which {\kdtrees} provide little computational saving because
almost all the points match the query and thus need to be visited.
In general this problem is unavoidable. But in one special case it {\em can}
be avoided---if we merely want to count the number of datapoints in a range
instead of explicitly find them all.

\begin{figure}[t]
\includegraphics[width=1.0\textwidth]{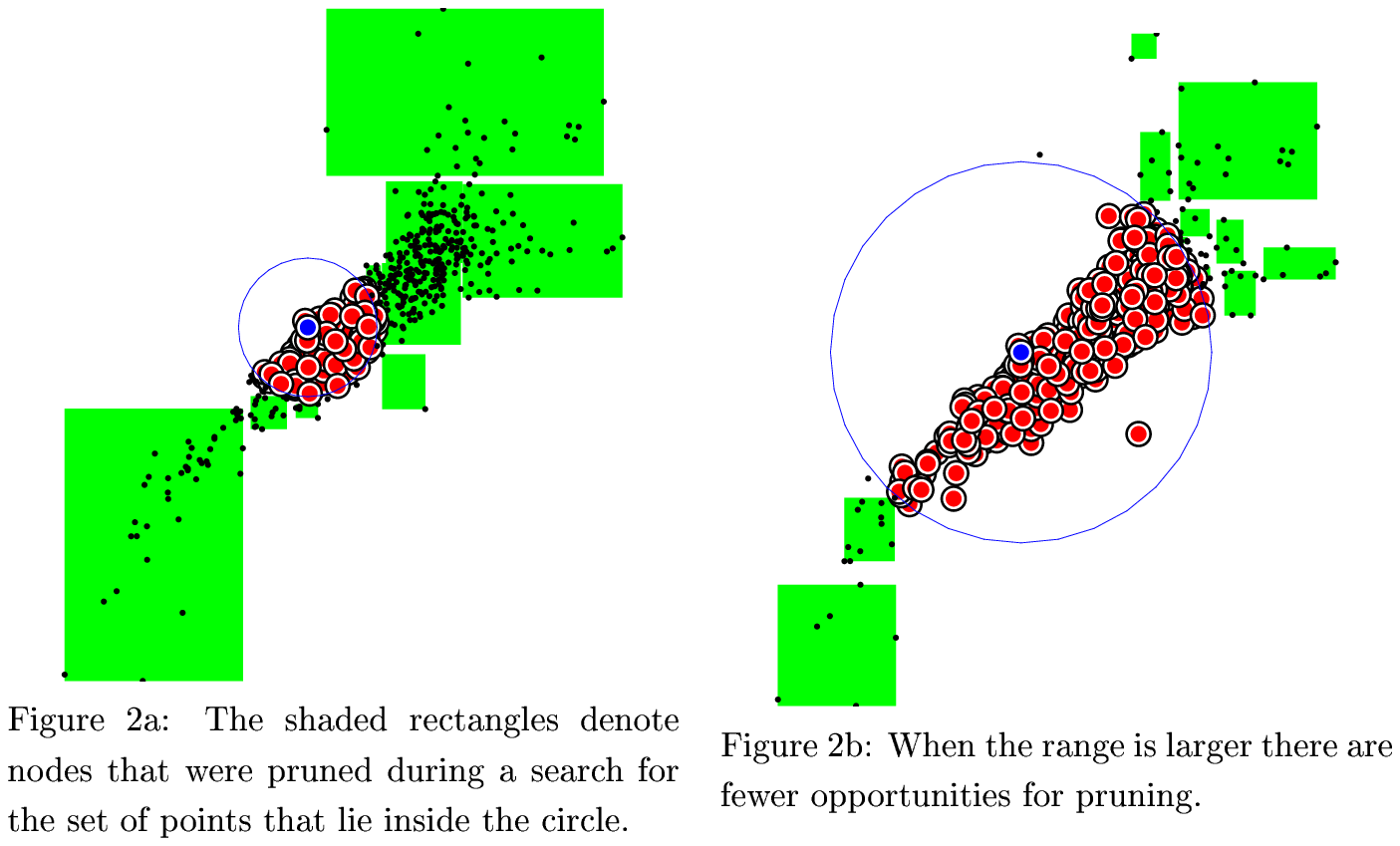}
\end{figure}

\subsection{Range Counting and Cached Sufficient Statistics}

We will add the following field to a {\kdtree} node. Let $\kdnodenumpoints$ be
the number of points contained in node $\kdnode$. This is the first and
simplest of a set of {\kdtree} decorations we refer to as {\em cached
sufficient statistics} (see Moore \& Lee 1998). In general, we frequently
stored the centroid of all points in a node and their covariance matrix.

Once we have $\kdnodenumpoints$ it is trivial to write an operation that
counts the number of datapoints within some range without explicitly visiting
them.

\begin{itemize}
\item
$\rangecount(\kdnode,\query,\hisep)$\\ 
Returns an integer: the number of
points that are both inside the $\kdnode$ and also within distance $\hisep$ of
$\query$.
\item
Let $\mindist$ := the closest distance from $\query$ to $\kdnodehr$.
\item
If $\mindist \geq \hisep$ then it is impossible that any point in
$\kdnode$ can be within range of the query. So simply return 0.
\item
Let $\maxdist$ := the furthest distance from $\query$ to $\kdnodehr$.
\item
If $\maxdist \leq \hisep$ then every point in $\kdnode$ must be within
range of the query. So simply return $\kdnodenumpoints$.
\item
Else, if $\kdnode$ is a leaf node, we must iterate through all the
datapoints in its leaf list. Start a counter at zero. For each point,
find if it is within distance $\hisep$ of $\query$. If so, increment
the counter by one. Return the count once the full list has been
scanned.
\item
Else, $\kdnode$ is not a leaf node. Then:
\begin{itemize}
\item
Let $\cleft := \rangecount(\kdnodeleft,query,\hisep)$
\item
Let $\cright := \rangecount(\kdnoderight,query,\hisep)$
\item
Return $\cleft + \cright$.
\end{itemize}
\end{itemize}

The same query that gave the poor range search performance in
Figure 2b gives good performance in
Figure 3. The difference is that a second type of
pruning of the search is possible: if the hyperrectangle surrounding
the {\kdnode} is either entirely outside {\em or inside} the range
then we prune.

\begin{figure}[t]
\setcounter{figure}{2}
\sidecaption\includegraphics[width=.7\textwidth]{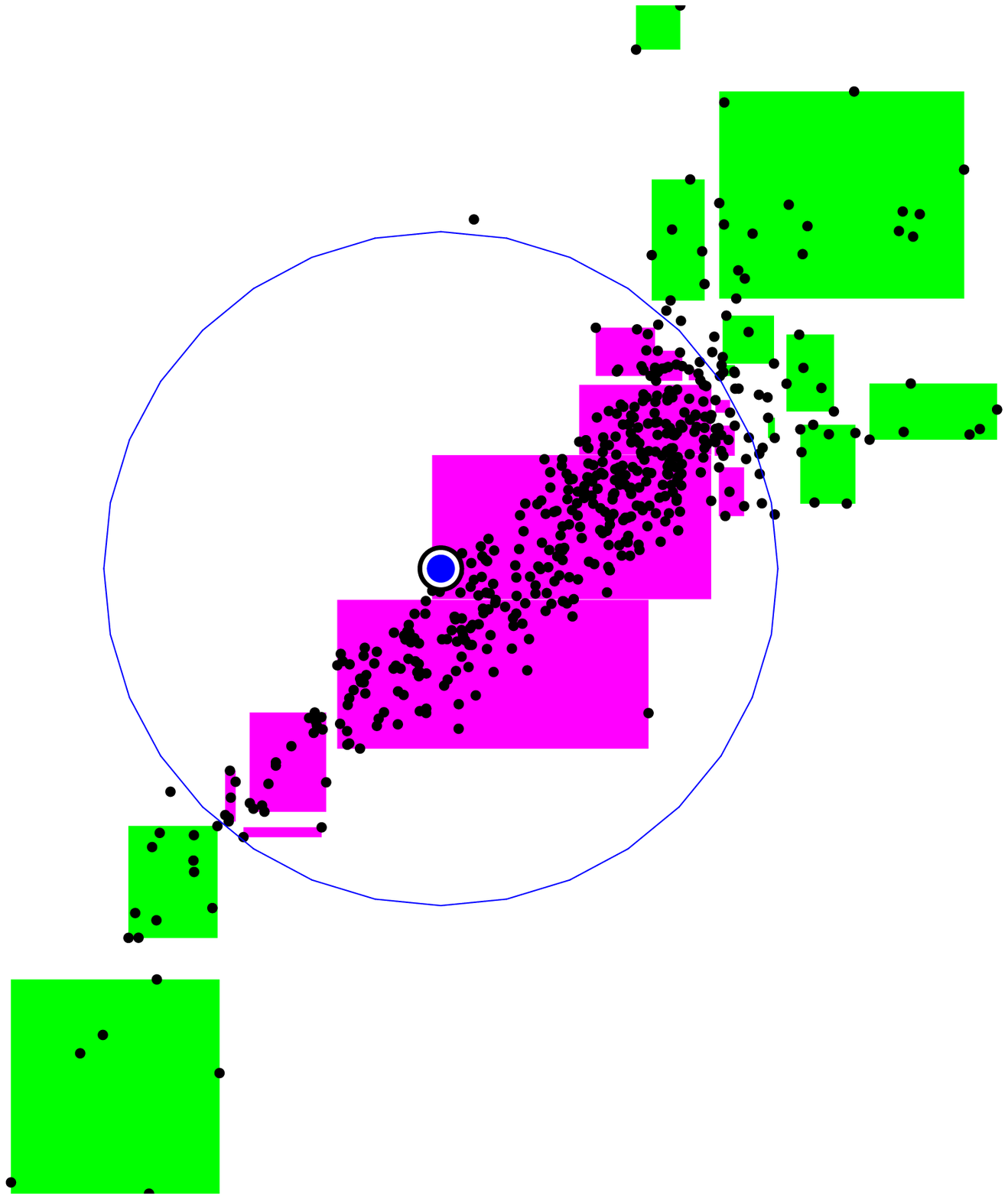}
\caption[width=.3\textwidth]{When doing a range count, nodes entirely within
range can also be pruned and added to the total count. This additional
pruning adds significant speed-ups to the slower range count discussed
in Figure 2b. Now one just spends time studying nodes on the boundary of
the range count.
\label{fig3}
}
\end{figure}

\section{Fast $N$--point Correlation Functions}
\label{npt}

\subsection{The Single Tree Approach to Two-Point Computation}

It is easy to see that the 2-point correlation function is simply a
repeated set of range counts.  For example, given a minimum
and maximum separation $\losep$ and $\hisep$ we run the following algorithm:

\begin{itemize}
\item
$\singletreetwopoint(\bigx,\kdnode,\losep,\hisep)$\\
Input $\bigx$ is a dataset, represented as a matrix in which the $k$th
row corresponds to the $k$th datapoint. $\bigx$ has $R$ rows and
$k$ columns. 
Input $\kdnode$ is the root of a kdtree built from the data in $\bigx$.
Output integer: the number of pairs of points $(\boldxi,\boldxj)$ such
that $\losep \leq | \boldxi - \boldxj | < \hisep$.
\item
{\countvar} := 0
\item
For $i$ between $1$ and $R$ do:
\begin{itemize}
\item
${\countvar} := {\countvar} + \rangecount(\kdnode,\boldxi,\hisep) - 
                 \rangecount(\kdnode,\boldxi,\losep)$
\end{itemize}
\end{itemize}
Note that in practice we do not use two range counts at each
iteration, but one slightly more complex rangecount operation
\begin{quote}
$\rangecountbetweenseps(\kdnode,\query,\losep,\hisep)$
\end{quote}
that directly counts the number of points whose distance from
$\query$ is between $\losep$ and $\hisep$.

\subsection{The Dual Tree Approach to Two-Point Computation}

The previous algorithm iterates over all datapoints, issuing a range
count operation for each. We can save further time by turning that
outer iteration into an additional kd-tree search.  The new search
will be a recursive procedure that takes two nodes, $\kdnodea$ and
$\kdnodeb$, as arguments. The goal will be to compute the number of
pairs of points $(\boldx,\boldy)$ such that $\boldx \in \kdnodea$,
$\boldy \in \kdnodeb$, and $\losep \leq | \boldx - \boldy | <
\hisep$.

\begin{itemize}
\item
$\dualtreecount(\kdnodea,\kdnodeb,\losep,\hisep)$\\
Returns an integer: the number of 
pairs of points $(\boldx,\boldy)$ such that $\boldx \in \kdnodea$,
$\boldy \in \kdnodeb$, and $\losep \leq | \boldx - \boldy | <
\hisep$.
\item
Let $\mindist$ := the closest distance between $\kdnodeahr$ and $\kdnodebhr$.
\item
If $\mindist \geq \hisep$ then it is impossible that any pair of points
can match. So simply return 0.
\item
Let $\maxdist$ := the furthest distance between $\kdnodeahr$ and $\kdnodebhr$.
\item
If $\maxdist < \losep$ then it is again impossible that any pair of points
can match. So simply return 0.
\item
If $\losep \leq \mindist \leq \maxdist < \hisep$ then 
all pairs of points must match. Use $\kdnodeanumpoints$ and $\kdnodebnumpoints$
to compute the number of resulting pairs
$\kdnodeanumpoints \times \kdnodebnumpoints$, and return that value.
\item
Else, if $\kdnodea$ and $\kdnodeb$ are both leaf nodes, 
we must iterate through all pairs of
datapoints in their leaf lists. Return the resulting (slowly computed)
count.
\item
Else at least one of the two nodes is a non-leaf. Pick the non-leaf
with the largest number of points (breaking ties arbitrarily),
and call it $\kdnodetosplit$. Call the other node $\kdnodenonsplit$. Then:
\begin{itemize}
\item
Let $\cleft := \dualtreecount(\kdnodesplitleft,\kdnodenonsplit,\losep,\hisep)$
\item
Let $\cright := \dualtreecount(\kdnodesplitright,\kdnodenonsplit,\losep,\hisep)$
\item
Return $\cleft + \cright$.
\end{itemize}
\end{itemize}
% There is an important subtelty in our action in the case that
% $\losep \leq \mindist \leq \maxdist < \hisep$. What should be the
% count of the number of matching pairs? In general is will simply
% be $\kdnodeacount \times \kdnodebcount$. But what if
% be $\kdnodea = \kdnodeb$? This happens, for example, at the root
% call of the recursive algorithm. In that case we should return
% \begin{equation}
% \mychoose{\kdnodeacount}{2}
% \end{equation}
% An even more sinister special case would occur if one node were
% a child of the other node. 
Computing a 2-point function on a dataset $\bigx$ then simply consists
of computing the value $C =
\dualtreecount(\knroot,\knroot,\losep,\hisep)$, where $\knroot$
is a
kd-tree built from $\bigx$, for a range of bins with 
minimum and maximum boundaries of $\losep$ and $hisep$.
We note here that the 2-point correlation function,
the quanity of interest is not simply $C$, but $C/2$ (the
number of unique pairs of objects).

A further speed--up can be obtained by simultaneously computing the
$\dualtreecount(\knroot,\knroot,\losep,\hisep)$ over a series of bins. We will
discuss this in further detail in Connolly et al.  (2001).

\subsection{Redundancy Elimination}

So far, we have discussed two operations -- exclusion and subsumption -- which
remove the need to traverse the whole tree thus speeding--up the computation
of the correlation function.  Another form of pruning is to eliminate
node-node comparisons which have been performed already in the reverse order.
This can be done simply by (virtually) ranking the datapoints
according to their position in a depth-first traversal of the tree, then
recording for each node the minimum and maximum ranks of the points it owns,
and pruning whenever $\kdnodea$'s maximum rank is less than $\kdnodeb$'s
minimum rank. This is useful for all-pairs problems, but will later be seen to
be {\em essential} for all-k-tuples problems. This kind of pruning is not
practical for Single-tree search.

\subsection{Multiple Trees Approach to $N$--Point Computation}

The advantages of Dual-Tree over Single-Tree are so far two
fold. First, Dual-tree can be faster, and second it can exploit
redundancy elimination.  But two more advantages remain. First, we can
extend the ``2-tree for 2-point'' method up to ``N-trees for
N-point''. Second (discussed in Section~\ref{se:approx}), we can
perform effective approximation with Dual-trees (or n-trees). We now
discuss the first of these advantages.

The $N$--point computation is parameterized by two $n \times n$ symmetric
matrices: $\lomat$ and $\himat$. We wish to compute
\begin{equation}
\sum_{i_1 = 1}^{R} 
\sum_{i_2 = 1}^{R} 
\ldots
\sum_{i_n = 1}^{R} 
\indicator{\lomat}{\himat}{i_1,i_2, \ldots i_n}
\end{equation}
where 
$\indicator{\lomat}{\himat}{i_1,i_2, \ldots i_n}$ is zero unless
the following conditions hold (in which case it takes the value 1):
\begin{equation}
\forall 1 \leq i < j \leq n ,
\lomatsub{i}{j} \leq |\boldxisub{i} - \boldxisub{j}| < \himatsub{i}{j}
\end{equation}
We will achieve this by calling a recursive function $\npt$ on an
$n$-tuple of kdtree nodes $(\kdnodesub{1},\kdnodesub{2} \ldots
\kdnodesub{n})$. This recursive function much return
\begin{equation}
\sum_{i_1 \in \kdnodesub{1}}
\sum_{i_2 \in \kdnodesub{2}}
\ldots
\sum_{i_n \in \kdnodesub{n}}
\indicator{\lomat}{\himat}{i_1,i_2, \ldots i_n}
\end{equation}

\begin{itemize}
\item
$\npt(\kdnodea,\kdnodeb,\losep,\hisep)$\\
\item
Let $\allsubsumed \assign \true$
\item
For $i = 1$ to $n$ do
\begin{itemize}
\item
For $j = i+1$ to $n$ do
\begin{itemize}
\item
Let $\mindist$ := the closest distance between $\kdnodeihr$ and $\kdnodejhr$.
\item
If $\mindist \geq \himatsub{i}{j}$ then it is impossible that any $n$-tuple of points can match because the distance between the $i$th and $j$th points in any such $n$-tuple must be out of range. So simply return 0.
\item
Let $\maxdist$ := the furthest distance between $\kdnodeihr$ and $\kdnodejhr$.
\item
If $\maxdist < \lomatsub{i}{j}$ then similarly return 0.
\item
If $\lomatsub{i}{j} \leq \mindist \leq \maxdist < \himatsub{i}{j}$ then 
every $n$--tuple has the property the the $i$th member and $j$th member 
match. We are interested in whether this is true for all $(i,j)$ pairs
and so the first time we are disappointed (by discovering the
above expression does not hold) then we will update the $\allsubsumed$ flag.
Thus the actual computation at this step is:
\begin{quote}
If $\lomatsub{i}{j} > \mindist$ or $\maxdist \geq \himatsub{i}{j}$ then \\ 
$\allsubsumed \assign \false$.
\end{quote}
\end{itemize}
\end{itemize}
\item
If $\allsubsumed$ has remained true throughout the above double loop,
we can be sure that every $n$--tuple derived from the nodes in the
recursive call must match, and so we can simply return
\begin{equation}
\prod_{i = 1}^{n} \kdnodeinumpoints
\end{equation}
\item
Else, if all of $\kdnodesub{1}, \kdnodesub{2}, \ldots \kdnodesub{n}$
are leaf nodes
we must iterate through all $n$--tuples of
datapoints in their leaf lists. Return the resulting (slowly computed)
count.
\item
Else at least one of the nodes is a non leaf. Pick the non-leaf
with the largest number of points (breaking ties arbitrarily),
and assume it has index $i = \istar$. Then:
\begin{itemize}
\item
Let 
$\cleft := \npt(\kdnodesub{1},\ldots,\kdnodeistarleft,\ldots,\kdnodesub{n})$
\item
Let 
$\cright := \npt(\kdnodesub{1},\ldots,\kdnodeistarright,\ldots,\kdnodesub{n})$
\item
Return $\cleft + \cright$.
\end{itemize}
\end{itemize}

The full $N$--point computation is achieved by calling $\npt$ with
arguments consisting of an $n$--tuple of copies of the root node.

We should note once again it is possible to save considerable
amounts of computation by eliminating redundancy. For example, in the
4-point statistic, the above implementation will recount each matching
4-tuple of points $(x,y,z,w)$ in 24 different ways: once for each of
the $4!$ permutations of $(x,y,z,w)$. Again, this excess cost can be
avoided by ordering the datapoints via a depth-first tree indexing
scheme and then pruning any $n$--tuple of nodes violating that order. But
the reader should be aware of an extremely messy problem regarding how
much to award to the count in the case that a subsume type of pruning
can take place. If all nodes own independent sets of points the answer
is simple: the product of the node counts.  If all nodes are the same
then the answer is again simple: $\mychoose{n}{N}$, where $n$ is the
number of points in the node. Somewhat more subtle combinatorics are
needed in the case where some nodes in the $n$-tuple are identical and
others are not. And fearsome computation is needed in the various
cases in which some nodes are descendants of some other nodes. 

\section{Controlled Approximation}
\label{apprx}

In general, when the final answer comes back from $\npt$, the majority of the
quantity in the count will be the sum of components arising from large subsume
prunes. But the majority of the computational effort will have been spent on
accounting for the vast number of small but unprunable combinations of
nodes. We can improve the running time of the algorithm by demanding that it
also prunes it search in cases in which only a tiny count of $n$--tuples is at
stake. This is achieved by adding a parameter, $\threshntuples$, to the $\npt$
algorithm, and adding the following lines at the start:
\begin{itemize}
\item
Let $\maxeffect \assign \prod_{i = 1}^{n} \kdnodeinumpoints$
\item
If $\maxeffect < \threshntuples$ then quit this recursive call.
\end{itemize}
This will clearly cause an inaccurate result, but fortunately it is
not hard to maintain tight lower and upper bounds on what the true
answer would have been if the approximation had not been made. Thus
$\npt(\kdnodesub{1},\kdnodesub{2},\ldots,\kdnodesub{n},\threshntuples)$
now returns a pair of counts $(\countlo,\counthi)$ where we can
guarantee that the true count $C$ lies in the range $\countlo \leq C
\leq \counthi$. 

\subsection{Iterative Deepening for Controlled Approximation}
\label{se:approx}

Suppose the true value of the $N$--point function is $C$ but that
we are prepared to accept a fractional error of $\epsilon$: we will be
happy with any value $\countapprox$ such that
\begin{equation}
\label{awareapprox}
|\countapprox - C| < \epsilon C
\end{equation}
It is possible to adapt the n-tree algorithm using a best-first
iterative deepening search strategy to guarantee this result while
exploiting permission to approximate effectively by building the count
as much as possible from ``easy-win'' node pairs while doing
approximation at hard deep node-pairs. This is simply achieved by
repeatedly calling the previous approximate algorithm with diminishing
values of $\threshntuples$ until a value is discovered that satisfies
Equation~\ref{awareapprox}.

\begin{figure}[t]
\sidecaption\includegraphics[width=.7\textwidth]{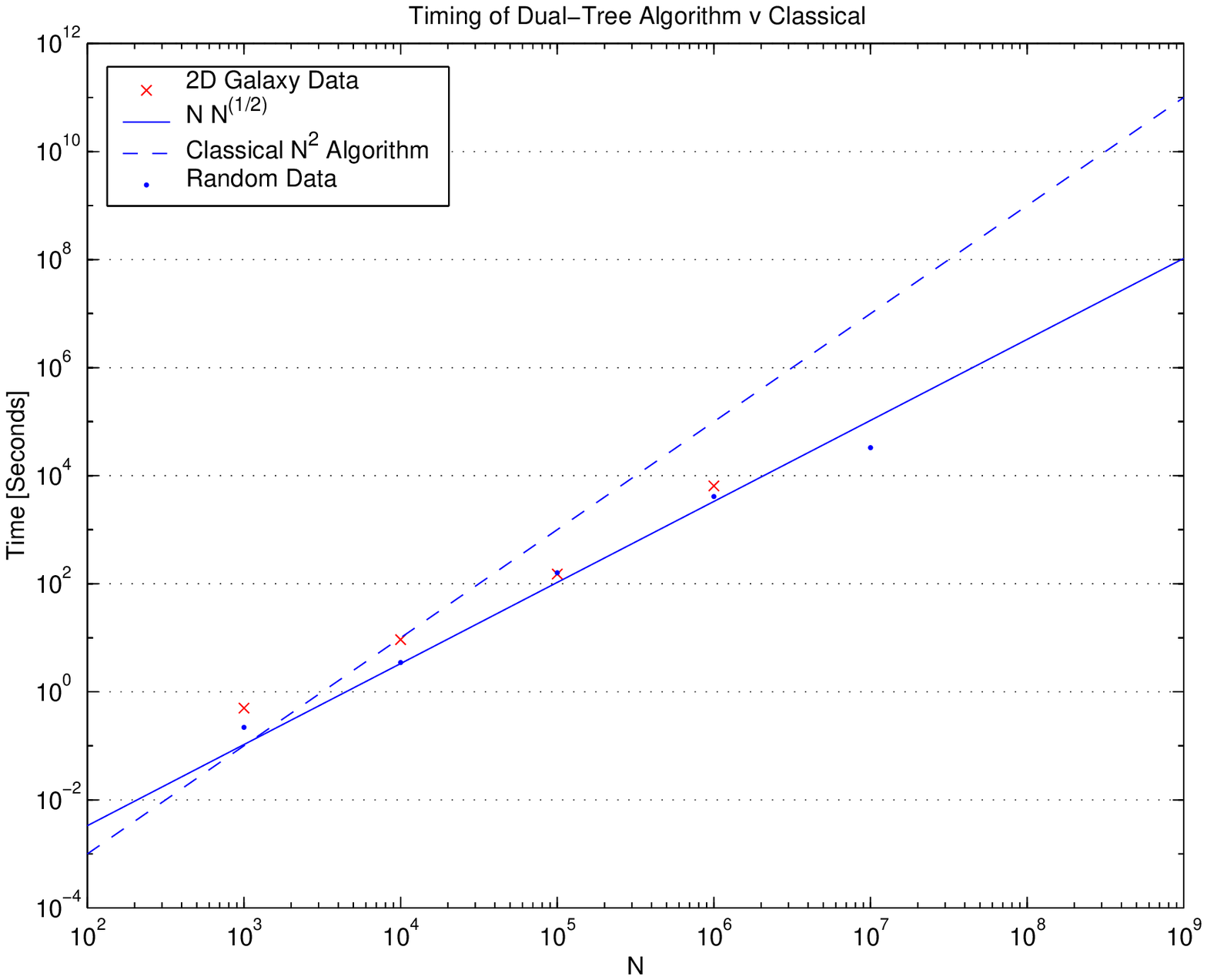}
\caption[width=.3\textwidth]{The computational time of our algorithm versus
the size of dataset. The crosses are real 2-dimensional projected galaxy data
while the dots are just drawn from a Poission distribution. The theoretically
expected scaling law of $N\sqrt{N}$ is shown and agrees well with the observed
data. The naive $N^2$ law is also plotted for comparison
\label{fig4}
}
\end{figure}

\section{Discussion}

We plan to present a more detailed discussion of the techniques presented here
in a forthcoming paper (Connolly et al. 2001). That paper will also include a
full analysis of the computational speed and overhead of our $N$--point
correlation function algorithm and compare those with existing software for
computing the higher--order correlation functions {\it e.g.} Szapudi et
al. 1999a. However, in Figure \ref{fig4}, we present preliminary results on
the scaling of computational timing needed for a 2-point correlation function
as a function of the number of objects in the data set. For these tests, we
computed all the data--data pairs for random data sets and real, projected
2-dimensional galaxy data. These data show that our 2-point correlation
function algorithm scales as $O(N\,\sqrt{N})$ (for projected 2-dimensional
data) compared to the naive all-pairs scaling of $O(N^2)$ where here $N$ is
the size of the dataset under consideration. To emphasis the speed--up
obtained by our algorithm (Figure \ref{fig4}), an all--pairs count for a
database of $10^7$ objects would take only 10 hours (on our DEC Alpha
workstation) using our methodology compared to $\sim10,000$ hours ($>1$ year)
using the naive $N^2$ method. Clearly, binning the data would also drastically
increase the speed of analyses over the naive all--pairs $O(N^2)$ scaling but
at the price of lossing of resolution.

Similar spectacular speed--ups will be achieved for the 3 and 4--point
functions and we will report these results elsewhere (Connolly et
al. 2001). Furthermore, controlled approximations can further accelerate the
computations by several orders of magnitude.  Such speed--ups are vital to
allow Monte Carlo estimates of the errors on these measurements. In summary,
our algorithm now makes it possible to compute an exact, all--pairs,
measurement of the 2, 3 and 4--point correlation functions for data sets like
the Sloan Digital Sky Survey (SDSS). These algorithms will also help in the
speed-up of Cosmic Microwave Background analyses as outlined in Szapudi et
al. (2000).

Finally, we note here that we have only touched upon one aspect of how trees
data structures (and other computer science techniques) can help in the
analysis of large astrophysical data sets. Moreover, there are other tree
structures beyond {\kdtrees} such as ball trees which could be used to
optimize our correlation function codes for higher dimensionality data.  We
will explore these issues in future papers.

%INDEX%%%%%%%%%%%%%%%%%%%%%%%%%%%%%%%%%%%%%%%%%%%%%%%%%%%%%%%%%%%%%%%
\clearpage
\addcontentsline{toc}{section}{Index}
\flushbottom
\printindex
%%%%%%%%%%%%%%%%%%%%%%%%%%%%%%%%%%%%%%%%%%%%%%%%%%%%%%%%%%%%%%%%%%%%%

\end{document}